\documentclass[useAMS,usenatbib]{mn2e}
\usepackage{epsf,amsfonts,amsmath,amssymb}
\usepackage{graphicx}
\usepackage{ulem}
\usepackage{mathrsfs}
\usepackage{color}
\usepackage{gensymb}

\title[Implications on X-ray flare models]{X-ray flares in GRBs: general considerations and photospheric origin}
\author[P. Beniamini $\&$ P. Kumar]{Paz Beniamini$^{1}$
\thanks{E-mail:paz.beniamini@gmail.com}, Pawan Kumar$^{2}$\\
$^{1}$Racah Institute for Physics, The Hebrew University, Jerusalem, 91904, Israel\\
$^{2}$University of Texas at Austin, Austin, TX 78712, USA}

\begin{document}

\date{Accepted ... Received ...; in original form ...}

\pagerange{\pageref{firstpage}--\pageref{lastpage}} \pubyear{2002}

\maketitle

\label{firstpage}

\begin{abstract}
Observations of X-ray flares from Gamma Ray Bursts (GRBs) imply strong constraints on possible physical models.
We provide a general discussion of these. In particular, we show that in order to account for the relatively flat and weak optical flux during the X-ray flares,
the size of the emitting region should be $\lesssim 3\times 10^{14}$cm. The bolometric luminosity of flares also strongly constrain the energy budget, and are inconsistent with late time activity of a central engine powered by the spin-down of a magnetar. We provide a simple toy model according to which flares are produced by an outflow of modest Lorentz factor (a few tens instead of hundreds) that is launched more or less simultaneously with the highly relativistic jet which produced the prompt gamma-ray emission.
The ``slower" moving outflow produces the flare as it reaches its photosphere.
If the X-ray flare jets are structured, the existence of such a component may naturally resolve the observational challenges imposed by flares, outlined in this work.
\end{abstract}

\section{Introduction}
\label{Int}
Early X-ray light-curves of Gamma Ray Bursts (GRBs) present complex temporal behaviour.
Specifically, in many GRBs the initial light-curve falls very steeply (this is known as the ``steep decay" phase) and
then exhibits a very shallow declining phase (``the plateau") before settling at the ``regular" ($F\sim t^{-1.5}$) decay expected from the external forward shock. One of the most interesting features in these light-curves is the occurrence of ``X-ray flares" in approximately 1/3 of GRBs \citep{Burrows(2005b),Falcone(2007),Chincarini(2010),Margutti(2011)}. These flares occur between $30-10^5$sec after the burst trigger, exhibit significant re-brightening (up to a factor of 500 in some cases) and are associated with the release of large isotropic equivalent energies (for the early time flares, on average 0.1 of the energy released during the prompt, and in some cases comparable). On average, these flares last $\sim 0.2t_f$ where $t_f$ is the time of the flare's onset, after which the light-curve goes back to the same temporal and spectral fluxes exhibited right before the flares. The existence of an underlying continuum with the same slopes before and
after the flaring activity as well as the flares' highly variable nature, makes it highly unlikely that flares are produced by the same component as the continuum emission dominating before and after each flare, i.e. the forward external shock. Therefore, the flares are likely the result of the same mechanism producing the prompt emission.

All this has led previous authors to consider the possibility that flares are produced by prolonged activity of the GRB central engine \citep{Burrows(2005a),FanWei(2005),Zhang(2006)}.
This activity, up to $10^5$sec after the trigger is non-trivial to understand in the context of black holes and provides some support for magnetar central engines for GRBs \citep{Kluzniak(1998),Dai(2006)}. Nonetheless, there have also been several proposals to generate flares within the black hole central engine models,\citep[see e.g. ][]{King(2005),Perna(2006),Proga(2006),Lee(2009),Cao(2014)}. In the present work we are not limiting ourselves to a specific central engine model, although some of the ``challenges" below, turn out to be more limiting for magnetars than for black holes.
\section{Motivation}
\label{motivation}
\begin{figure*}

\centering
\includegraphics[width=0.8\textwidth]{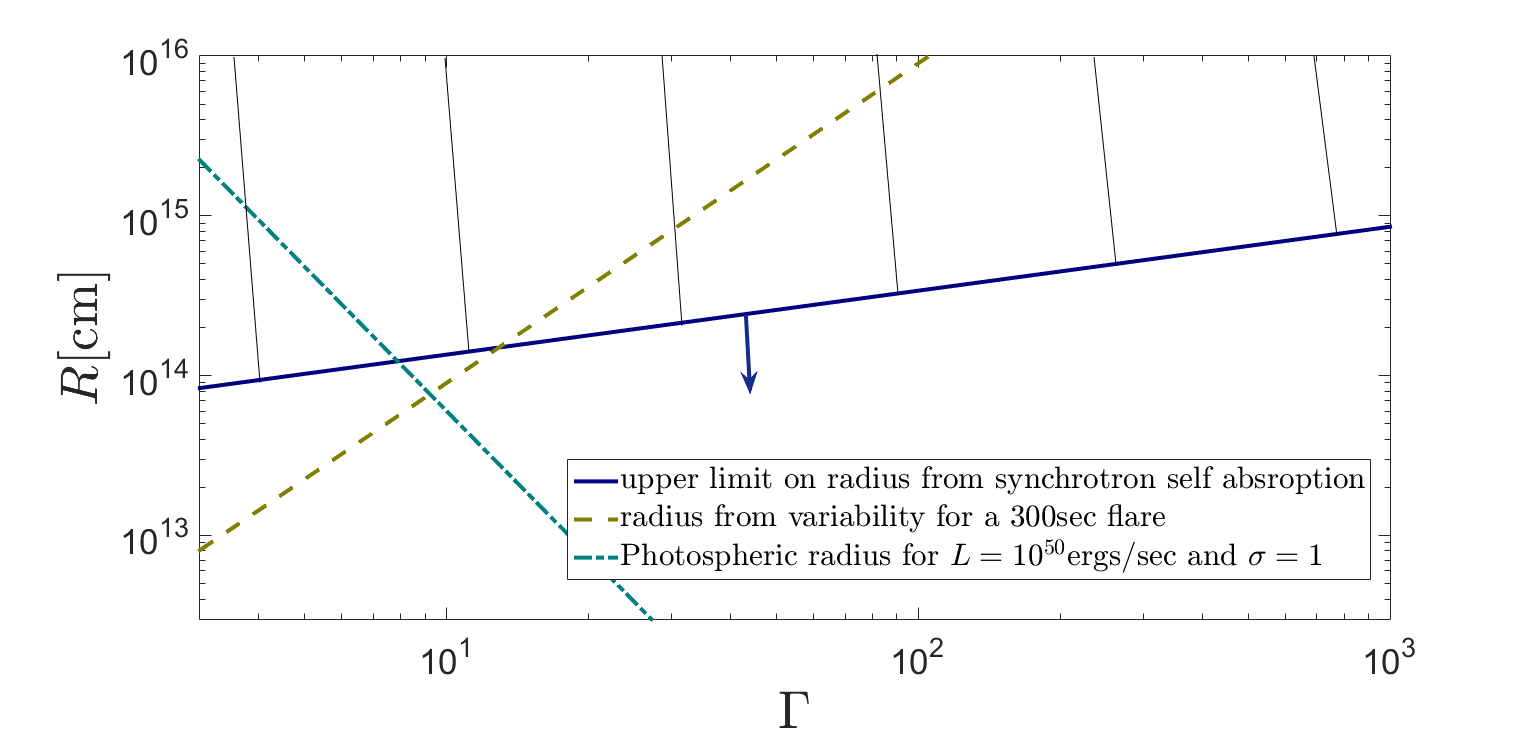}
\caption
{\small Upper limit on the radius assuming synchrotron emission and requiring that the optical flux is sufficiently suppressed below the observed limits (solid line). This upper limit may be smaller for Baryonic jets (see \S \ref{motivation} for details).
In comparison we plot also the radius given by the variability time-scale for a flare starting at 300sec with a variability $\Delta t\approx 60$sec (dashed line), and the photospheric radius (considered to produce the flares in this paper) assuming $L_{f}=10^{50}$ergs/sec and $\sigma = 1$ (dot dashed line). In many models (including the one considered in this paper) the second of these radii determines the radius of emission for the flares of given variability. In such cases the Lorentz factor of the flare emitting material is significantly limited to $\Gamma \lesssim 20$ (for flares starting at 300sec). In addition we see from the intersection of the dashed and dot-dashed lines that the radius considered in this paper is consistent with the limits from synchrotron self-absorption even for $\sigma = 1$ (for larger $\sigma$ the radius decreases as $R \propto (1+\sigma)^{-1}$).}
\label{radiuslim}
\end{figure*}

We outline a number of challenges that must be resolved in any model that attempts to explain X-ray flares.
\begin{enumerate}
\item There is a broad range of typical flare time-scales. In addition, these times are generally much longer than the prompt variability time-scale. Any model attempting to explain flares, should be able to reproduce this range of time-scales. This is non-trivial; for instance, in the magnetar scenario the natural time-scale would be the magnetar spin down time. If this time is associated with either the prompt or the flares, it would not be able to account for the other.
\item While strong flares are regularly observed in X-rays, no strong flaring behaviour is observed in the optical band at the same
time \citep[e.g.,][]{Swenson(2013)}. This implies that the spectral slope of the flare emission between these two bands must be very hard.
Upper limits on the optical flux at a few tens up to a few hundreds of seconds, are typically between 0.1mJy and 5mJy (at $\sim 2eV$) \citep{Santana(2015),Troja(2015)}.
Specifically, \cite{Santana(2015)}, study 8 flares selected for their brightness in X-rays, showing that in all these flares the spectral slopes between the optical and X-ray bands, should be harder than $F_{\nu}\sim \nu^{0.7}$ in order not to over produce optical radiation as compared with observations.
This significantly limits possible emission models for the flares, and in particular implies that in order for synchrotron models
(either in the slow or fast cooling regimes) to be viable, the spectrum has to become self absorbed somewhere between optical and X-rays, and the spectral slope below this frequency should become much harder.
A similar situation occurs during the prompt phase and has been used to significantly limit the possible emission mechanism of the prompt GRB and the jet composition at the emission site \citep{Shen(2009),Beniamini(2014),Kumar(2014)}.

For the case of slow cooling synchrotron,
even a self absorption break at $\nu_0\sim 3 \nu_{opt}$ (where $\nu_{opt}$ is the frequency of the optical band) is typically sufficient to suppress the optical flux below the observed limits during the flares starting at $ t_f\approx 300$sec
We note that in the fast cooling regime,
the extrapolated flux below the X-ray band increases and therefore also the required $\nu_0$ to sufficiently suppress the optical flux, increases.
However, by definition, the quantity: $F_0/\nu_0^{\alpha}$ (where $\alpha$ is the spectral slope below the self absorption frequency)
will remain constant in this process, thus leaving the upper limits on the radius given by Eq. \ref{eq:radiuslim} unchanged.
For the synchrotron self absorption frequency, $\nu_{SA}$, to be larger than $\nu_0$, the maximum allowed radius where the radiation is produces is given by:
\begin{equation}
\frac{2\nu_{SA}^2}{c^2}\gamma_e(\nu_{SA}) \Gamma m_e c^2 \frac{R^2}{4\Gamma^2 d_L^2}\leq F_{obs}(\nu_X)\bigg(\frac{\nu_{SA}}{\nu_X} \bigg)^{1/3}
\end{equation}
where $\Gamma$ is the Lorentz factor of the matter producing the flare, $F_{obs}(\nu_X)$ is the observed specific flux at $\nu_X=2keV$ during the 
flares starting at $300$sec \citep{Santana(2015)}, $d_L$ is the luminosity distance and $\gamma_e(\nu_{SA})$ is the typical thermal Lorentz factors of the electrons radiating synchrotron at $\nu_{SA}$ and is given by:
\begin{equation}
\label{gammae}
\gamma_e(\nu_{SA})=\bigg(\frac{2 \pi m_e c \nu_{SA}}{\Gamma q B'}\bigg)^{1/2}=\bigg(\frac{\sqrt{2} \pi m_e c^{3/2} \nu_{SA}\sqrt{1+\sigma}R}{q \sqrt{L}\sqrt{\sigma}}\bigg)^{1/2}
\end{equation}
where $B'$ is the magnetic field in the jet frame and in the last transition it has been related to the jet luminosity, $L$, the radius, $R$ and the magnetization $\sigma$ (the relative energy density in magnetic fields compared to that in particles).
This leads to:
\begin{equation}
\label{eq:radiuslim}
R<3.4\times 10^{14} \bigg(\frac{F_{obs}(\nu_X)}{10 mJy}\bigg)^{0.4}\!\bigg(\frac{\Gamma}{100}\bigg)^{0.4}d_{L,28}^{0.8}L_{50}^{0.1}\bigg(\frac{\sigma}{1+\sigma}\bigg)^{0.1}cm
\end{equation}
where $L_{50}=L/10^{50}\mbox{ergs/sec}$ and $d_{L,28}=d_L/10^{28}$cm. This consideration therefore significantly limits the allowed radius
for producing the flares.
This upper limit can become even smaller in case particles are accelerated in shocks (which is likely for Baryonic jets) to form a non-thermal spectrum above some minimum Lorentz factor $\gamma_m$. In this case $\gamma_e(\nu_{SA})$ in Eq. \ref{gammae} is replaced by $max(\gamma_e(\nu_{SA}),\gamma_m)$ and the limit on $R$ could be decreased. In addition, PIC simulations suggest that locally the magnetic field may be significantly weaker than the equipartition value for shock acceleration of particles \citep{Sironi(2011)}. This would require slightly hotter electrons (since from Eq. \ref{gammae}, $\gamma_e(\nu_{SA}) \propto \sigma^{-1/4}$) and would imply smaller radii given the same observed flux.
The limits on radius and Lorentz factor are shown in Fig. \ref{radiuslim}, in comparison with the radius determined by the variability time scale for a flare starting at $300$sec. In case the emission radius is set by the latter, the Lorentz factor of the flare producing material is also significantly constrained by this consideration, to be $\lesssim 20$. 
%In addition, for some flares, such as in GRB 051117A the observed fluxes lead to even stronger upper limits on the radius, smaller than $3\times 10^{14}cm$.
\item In the magnetar model, the energy source for the flares would be the rotational energy of the magnetar. However, the magnetar is expected to spin down very fast, leading to a decrease in the available energy for flares (which would become more severe for flares emitted at later times). For a general spin-down mechanism with a braking index of $n$, the energy goes down as $E \sim t^{2 \over 1-n}$, which for dipole radiation implies $E\sim t^{-1}$. 
The energy in flares, indeed seems to be falling down in time (at least initially), although at a somewhat different rate: \cite{Margutti(2011)} show that $E \sim t^{-1.7}$ for flares that start up to $\sim 1000$sec and $E\sim const$ for flares between 1000sec and $10^5$sec.
However, the real problem is that the breaking index ($n$) should be $\sim 1.7$ in order for the magnetar model to be able to explain the rapid decline phase of the X-ray light-curves \citep{Tagliaferri(2005)}, seen between $\sim 10^2-10^3$sec. 
In other words, one needs $E \propto t^{-3}$ or faster to explain the X-ray steep decline up to $\sim 1000$sec after the trigger, but in that case the rotational energy in the magnetar is too little to explain the late time flares.

\item A related issue has to do with the energy required for the jet to carve out a cavity through the polar region of the GRB progenitor star. The energy required for this process is likely very large (e.g \citealt{Woosley(1993),MacFadyen(1999),Ramirez-Ruiz(2002),Matzner(2003),Bromberg(2015)}). Therefore, if the central engine stops operating after producing the prompt gamma-rays and then restarts at $\sim 300$sec or later, the polar cavity opened by the prompt jet would have already closed and the flare material would have to re-open this cavity. It would therefore have to be initially significantly more energetic as compared with estimates based on the flux eventually emitted during the flares (which are already huge and highly constraining, as mentioned above). 
\end{enumerate}

\section{The Basic Model}
\label{Basis}
Motivated by the considerations discussed in \S \ref{motivation}, in particular X-rays being produced close to the photosphere and the difficulty of reopening the polar cavity if the X-ray jet is launched after a period of central engine inactivity, we present a model for X-ray flares where the central engine activity is confined to a short time period of order less than $10^2$sec.
We consider a scenario in which the prompt emission and flare(s) are emitted from the photospheric radius. In our model, both the material producing the prompt
phase of the emission and that producing the flare(s) are ejected from the central engine at the same time, but with different velocities. The fast material produces the prompt phase of the emission as it reaches its photosphere, whereas the slower material eventually dissipates its energy and produces the flare(s) as it reaches its own (further out) photosphere.
Due to the steep dependence of the photosphere (and the corresponding observed time-scale for the emission) on the Lorentz factor of the material, a variation in the Lorentz factor by less than an order of magnitude is likely to be sufficient to account for the large difference in the prompt vs. flare typical variability time-scales.
This model may naturally resolve the challenges presented above associated with producing flares:
\begin{enumerate}
\item The different time-scales associated with the onset and duration of the prompt and flares, are naturally reproduced from a single time-scale at the source.
\item The small radius from which the flares are emitted enables the optical radiation to be sufficiently suppressed below detection level, consistent with observations.
\item Since the flare emitting material is ejected together with or just after the prompt emitting material, there is sufficient energy in the magnetar at the time of ejection to produce the flare.
\item The fact that the flare emitting material is ejected together with the prompt emitting material, implies also that the former can pass more easily through the surrounding environment, and does not have to invest large amounts of energy just to re-open the cavity in order to punch through the surface of the progenitor star, as is the case for models invoking late time activity for the central engine.
\end{enumerate}

\section{Emission radii, time-scales and spectra}
Consider a jet moving at a Lorentz factor of $\Gamma$. The jet luminosity is given by:
\begin{equation}
\label{Ljet}
L=(1+\sigma)\Gamma\dot{M}c^2
\end{equation} 
where $\dot{M}$ is the mass loss rate.
The optical depth to Compton scatterings is given by:
\begin{equation}
\tau\approx n'\sigma_T \frac{R}{\Gamma}=\frac{\dot{M}}{4\pi \Gamma R^2 c m_p}\frac{\sigma_T R}{\Gamma}=\frac{L \sigma_T}{(1+\sigma)4\pi R c^3 m_p \Gamma^3}
\end{equation}
where $n'$ is the jet density in the co-moving frame and where we have made use of Eq. \ref{Ljet}.
The optical depth decreases with increasing radius until it reaches $\tau=1$ at the photosphere, which we assume as the radius of emission (see Fig. \ref{radiuslim}). The emission radius in this model may be slightly below the photosphere, as in any case the radiation will be trapped up until the photospheric radius. However, dissipation cannot occur much below this radius, since in this case the jet could suffer significant adiabatic losses between the dissipation and emission radii, leading to greatly increased requirements on the energy source and resulting in a significant re-brightening of the afterglow, contrary to observations.
Assuming that the shell contributing to the radiation is narrow ($\Delta R< R/\Gamma^2$), emission from this radius would last for:
\begin{equation}
\label{tphoto}
\Delta t=\frac{R}{2c\Gamma^2}=\frac{L \sigma_T}{(1+\sigma)8\pi  c^4 m_p \Gamma^5}.
\end{equation}
Notice that this time depends very strongly on $\Gamma$, implying a large range of variability times, from the shortest flares to the longest ones, would be obtained from a relatively narrow distribution in $\Gamma$ (this would also imply the same underlying mechanism producing the prompt and flares' variabilities). We note that it is very natural to expect a distribution of Lorentz factors in the flow, since it is unlikely that the flow would be very regular.
Assuming that both the prompt phase and the flares are produced from their corresponding photospheres, Eq. \ref{tphoto}, implies that:
\begin{equation}
\frac{\Delta t_{flare}}{\Delta t_{GRB}}=\bigg(\frac{L_{f}}{L_{GRB}}\bigg)\bigg(\frac{1+\sigma_{GRB}}{1+\sigma_{f}}\bigg)\bigg(\frac{\Gamma_{GRB}}{\Gamma_{f}}\bigg)^5,
\end{equation} 
where the sub-script ``GRB" refers to the parameters for the material creating the prompt emission and the sub-script ``f" to the same parameters for the flare emitting material.
As an example, for a flare starting at $\sim 300$sec, with $\Delta t_{flare}\approx 60$sec, $L_f\approx 0.01 L_{GRB}$ a decrease in $\Gamma$ by a factor of $\sim 6$ (for the same magnetization) is sufficient to obtain a $\Delta t_{flare}\approx 60\Delta t_{GRB}\approx 60$sec.
Due to the strong dependence on $\Gamma$, the qualitative result will hold even if the magnetization is quite different for the different materials. In addition, since the flare emitting material was emitted at early times,
roughly together with the prompt emitting material, a magnetar source would still have sufficient rotational energy during the launching phase
to power the observed flare energies (see \S \ref{motivation}).

Finally, consider the spectrum below the X-ray band. 
Since we are considering photospheric emission, the emitting region is compact and the spectrum in our model is likely to be self absorbed above the optical band. Keeping the emission mechanism general, $\nu_{SA}$, satisfies:
\begin{equation}
\frac{2\nu_{SA}^2}{c^2}\gamma_e \Gamma m_e c^2 \frac{R^2}{4\Gamma^2 d_L^2}=F_{obs}(\nu_{SA})
\end{equation}
where $\gamma_e$ is the typical thermal Lorentz factor of the emitting electrons and $F_{obs}(\nu_{SA})$ is the observed flux at $\nu_{SA}$.
The largest possible radius where the emission can be produced and still become self-absorbed in the optical band is the photospheric radius. Plugging this into the equation above implies:
\begin{equation}
\label{SelfAbs}
\nu_{SA}\!\geq\! 2\times 10^{15}\bigg(\frac{F_{obs}(\nu_{SA})}{mJy}\bigg)^{0.5}\!(1+\sigma_{f})^{0.3} L_{f,50}^{-0.3}\gamma_e^{-0.5}d_{L,28} Hz
\end{equation} 
where we have used $\Delta t_{flare}\approx 60$sec typical for a flare starting at 300sec to obtain a lower limit on $\Gamma_f$. Given the large expected value of $\sigma_f$, Eq. \ref{SelfAbs} implies that the
spectrum is very likely self absorbed above the optical band (as required by observations) even if the electrons producing the radiation have quite large $\gamma_e$. Specifically, we note that this holds for the case of synchrotron emission, as can be seen from Eq. \ref{eq:radiuslim} and Fig. \ref{radiuslim}. Note however, that though necessary, this consideration is not sufficient. It is possible that even if the radiation is produced at $\tau\geq 1$, the spectrum below $\nu_{SA}$ can be flat due to the effect of IC scatterings
(\citealt{Thompson(1994),Ghisellini(1999),Meszaros(2000)}, etc.).
This condition should therefore be self-consistently verified for any detailed model attempting to explain GRB flares.
\section{Implications on the distribution of velocities in the jet}
Given the distribution of flare luminosity with time, we can determine the required distribution of jet luminosity as a function of $\Gamma$, within the model proposed in this paper. The observed distribution is approximately \citep{Margutti(2011)}:
\begin{equation}
L(t)\propto
\left\{ \!
  \begin{array}{l}
t^{-2.7}\quad t\lesssim 1000\mbox{ sec}\\
 \\
 t^{-1} \quad t \gtrsim 1000 \mbox{ sec}\\
  \end{array} \right.
\end{equation}
In our model, we have $L(t) \propto t(1+\sigma)\Gamma^5$, leading to:
\begin{equation}
L(\Gamma)\propto
\left\{ \!
  \begin{array}{l}
(1+\sigma)^{0.73} \Gamma ^{3.65} \quad \Gamma \gtrsim \Gamma_{min}\\
 \\
 (1+\sigma)^{0.5} \Gamma ^{2.5} \quad \Gamma \lesssim \Gamma_{min}\\
  \end{array} \right.
\end{equation}
\section{Conclusions}
Observations of GRB flares strongly constrain the possible models. Specifically, models should be able to explain the broad range of typical observed time-scales and the strong suppression of the optical compared to the X-ray flux. 
Although a synchrotron origin for the observed X-ray emission cannot be ruled out by the data, the relative lack of optical to X-ray flux
implies a self absorbed source below $\sim 10eV$, leading to a small emission radius: $R\lesssim 3\times 10^{14}cm$ and also a relatively small Lorentz factor, in case the radius is set by the variability time-scale ($\Gamma\lesssim 20$ for a flare starting at $\sim 300$sec).
In addition flare models should be able to account for the considerable amounts of energies to be released as late flares (which poses a difficulty for the late time central engine activity model that invokes a rapidly rotating magnetar) and have enough energy not only to power the observed flare but also to allow the flare emitting jet (or outflow) to punch through the progenitor star.

We have considered a model in which X-ray flares and prompt GRB radiation have a common central engine.
We suggested that both the material producing the prompt GRB and that producing the flare(s) are ejected from the central engine at more or less the same time but at different speeds. The fast outflow produces the prompt GRB as it reaches its photosphere, whereas the slower material eventually produces the flare(s) as it reaches its own (further out) photosphere.
Due to the steep dependence of the observed time-scale in this model on the Lorentz factor of the material, a small variation in the Lorentz factor can account for the large difference in the prompt vs. flare typical variability time-scales.
The small emission radius in this model can allow for self absorbed spectra at the optical band. In addition, since the flare jet is produced at an early time, there is still a large energy reservoir available at that time. Finally the flare material follows on the tail end of the prompt jet, before the polar cavity of the star has closed, and so can break out through the surface of the GRB progenitor star with little expenditure of energy.

Another important test for any model attempting to explain late time flares regards their observed steep decays.
In models in which the flare variability is associated with the angular time-scale, one can expect the decay to be dominated by high latitude emission ($F_{\nu}\propto \nu^{-\beta} t^{-2-\beta}$ where $\beta$ is the spectral slope). However, observationally, the decays are much steeper, with decay slopes between -10 and -100 even after correcting for the most conservative value of the ``zero time", $t_0$ \citep{Uhm(2015)}.
One of the ways that a very fast decay (including, potentially, an exponential decay) of the X-ray light-curve may arise in the model proposed here, is if the angular size of the X-ray flare jet is $\lesssim 1/\Gamma$. This is natural to expect in our model since the flare jet Lorentz factor is of order $\sim 10$ (Fig. \ref{radiuslim}) and its angular size is set by the size of the polar cavity in the progenitor star that the high Lorentz factor (and high luminosity) gamma-ray producing jet had carved out (which we know from observations to be of order $5\degree-10\degree$). Since in this case there is very little radiation produced at latitudes $>1/\Gamma$, the light-curve's slope due to high-latitude emission is not limited to $2+\beta$, and in fact, the flux could fall off much more rapidly.
The exact shape of the light-curve will depend on the angular spread of energies within the flare producing jet, $dE/d\Omega$, on the Lorentz factor at different directions, $\Gamma(\theta)$ and on the rate of jet energy dissipation below the Thomson photosphere at different radii and angles. A further test for the applicability of such models would be to perform a detailed calculation of the expected light-curves from these models and compare the results against the observed light-curves.

Finally, we remark that although they have different energetics and time-scales, flares share many similarities with prompt GRB pulses, such as ``fast rise - exponential decay" time profiles \citep{Curran(2008)}, a pulse width that decreases with the observed frequency as approximately $W \sim \nu^{-0.5}$ \citep{Chincarini(2010)} and a correlation between peak frequency and luminosity \citep{Margutti(2010)}. This suggests that either these effects are purely dynamical \citep[see e.g.][]{BG(2015)} or else GRB pulses and flares share a common radiation mechanism. A better understanding of GRB flares could shed some light on the yet unknown prompt emission mechanism.
\section*{Acknowledgements}
We would like to thank Raffaella Margutti, Rodolfo Barniol Duran and Patrick Crumley for helpful discussions.
We are very grateful to the referee for his/her careful reading of the paper, and for extremely insightful comments and suggestions which substantially improved the paper.

\end{document}